\documentclass[journal,9pt]{IEEEtran}
\ifCLASSINFOpdf

\else
\fi
\usepackage{algorithm}
\usepackage{algorithmic}

\usepackage[colorlinks,linkcolor=blue,anchorcolor=black,citecolor=black]{hyperref}

\usepackage{mathrsfs}
\usepackage{amsfonts,amssymb,amsthm,latexsym}
\usepackage{amsmath}
\usepackage{xcolor}
\usepackage{cite}
\usepackage{array}
\usepackage{cases}
\usepackage{graphicx}
\usepackage{subfigure}
\usepackage{tikz}
\usetikzlibrary{arrows.meta}

\newtheorem{problem}{Problem}
\newtheorem{thm}{Theorem}
\newtheorem{lem}{Lemma}
\newtheorem{rem}{Remark}

\newtheorem{proposition}{Proposition}
\newtheorem{corollary}{Corollary}

\begin{document}
    \title{Almost-Global Planar Formation Control on Any Connected Graph Without a Common Orientation
   \author{Jingyi Zhao, Yuhu Wu, H\'{e}ctor Garc\'{i}a de Marina$^*$}
\thanks{This work was supported by the ERC Starting Grant iSwarm 101076091.}
\thanks{Jingyi Zhao and H\'{e}ctor Garc\'{i}a de Marina (corresponding author) are with Department of Computer Engineering, Automation, and Robotics, and with IMAG, University of Granada, 18001, Granada, Spain (e-mail: jingyizhao@ugr.es; hgdemarina@ugr.es) }
\thanks{Yuhu Wu is with the Key Laboratory of Intelligent Control and Optimization for Industrial Equipment of Ministry of Education and the School of Control Science and Engineering, Dalian University of Technology, 116024, Dalian, China (e-mail:wuyuhu@dlut.edu.cn).}
}

\maketitle

\begin{abstract}
	Formation control based on geometric constraints often requires either rigid interaction graphs or a common orientation reference. This paper shows that a prescribed planar formation can be achieved, up to translation and rotation, using the relative displacements measured over the $N-1$ edges of any tree interaction graph, together with one scalar per edge direction exchanged between neighbors. Since a tree is the sparsest connected graph, any connected set of available sensing links suffices. Each \emph{internal} agent of the tree parameterizes the positions of its tree neighbors on circles of the prescribed radii by means of virtual phases, and the relative phases within each neighborhood encode the internal geometry of the formation without fixing its orientation. The resulting distributed controller is a gradient flow. We characterize its complete equilibrium set: besides the desired formation, the only equilibria correspond to the collapse of a subset of \emph{internal tree edges}, and every such equilibrium has exactly as many unstable directions as collapsed edges. Consequently, the desired formation is locally exponentially stable and almost globally attractive, and it is globally attractive when the tree is a star. The controller is implemented in mutually misaligned local frames, which only need to share the same handedness, i.e., the same sense of rotation.
\end{abstract}


\IEEEpeerreviewmaketitle

\section{Introduction}
Formation control focuses on coordinating multiple agents to achieve and maintain a prescribed geometric configuration \cite{oh2015survey}. Depending on the geometric constraints used to specify the target formation, existing approaches are broadly categorized into distance- \cite{demarina2016distributed,krick2009stabilisation}, displacement- \cite{de2020maneuvering,ren2007consensus}, bearing- \cite{cheng2024general,su2026bearing}, and angle-based methods \cite{chen2020anglesss,li2022satellite}. These geometric representations differ in requirements for inter-agent information and interaction topology, as well as in the degrees of freedom retained by the prescribed formation.

\begin{figure}
	\centering
	\begin{tikzpicture}[scale=1.15, every node/.style={font=\footnotesize},
		ag/.style={circle,draw,fill=white,minimum size=5.5mm,inner sep=0pt},
		link/.style={blue!70!black,line width=1.4pt},
		blocked/.style={red!75!black,dashed,line width=1pt},
		obstacle/.pic={
			\filldraw[fill=brown!40,draw=brown!65!black,line width=0.6pt] plot[smooth cycle, tension=0.7] coordinates
				{(1.42,1.15) (1.25,1.5) (0.85,1.55) (0.5,1.3) (0.45,0.8) (0.7,0.48) (1.2,0.47) (1.5,0.75)};
			\node[brown!35!black,font=\scriptsize] at (1,1) {obstacle};}]
		\begin{scope}
			\coordinate (a1) at (0,2); \coordinate (a2) at (2,2); \coordinate (a3) at (0,0); \coordinate (a4) at (2,0);
			\draw[blocked] (a1) -- (a4);
			\draw[link] (a1) -- (a2) -- (a4) -- (a3) -- cycle;
			\pic {obstacle};
			\node[red!75!black,font=\Large] at (0.42,1.58) {$\times$};
			\foreach \i in {1,...,4}{\node[ag] at (a\i) {\i};}
			\node at (1,2.55) {Rigid graph: $2N-3=5$ links};
		\end{scope}
		\begin{scope}[xshift=3.3cm]
			\coordinate (a1) at (0,2); \coordinate (a2) at (2,2); \coordinate (a3) at (0,0); \coordinate (a4) at (2,0);
			\draw[link] (a3) -- (a1) -- (a2) -- (a4);
			\pic {obstacle};
			\foreach \i in {1,...,4}{\node[ag] at (a\i) {\i};}
			\node at (1,2.55) {Tree: $N-1=3$ links};
		\end{scope}
	\end{tikzpicture}
	\caption{Formation control around an obstacle with $N=4$ agents. Left: a minimally rigid distance graph needs $2N-3=5$ links; besides the perimeter, it needs a diagonal (dashed), which the obstacle blocks. Right: the proposed method only needs a tree with $N-1=3$ links, e.g., the path $3$--$1$--$2$--$4$.}
	\label{fig4}
\end{figure}
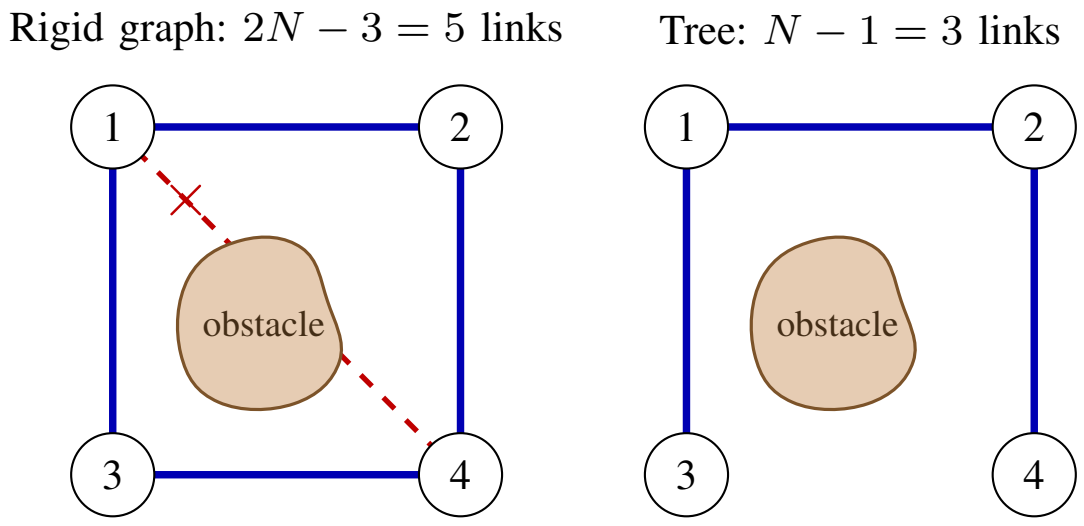

The interaction topology is particularly relevant when only limited relative measurements are available. For example, distance-based formation control typically requires a rigid interaction graph to locally determine the target formation up to rigid-body transformations, with a minimally rigid graph containing $2N-3$ edges in 2-D. However, relative measurements obtained through onboard sensors, such as cameras and LiDAR, may be restricted by Line-of-Sight (LoS) obstructions, preventing the available sensing links from satisfying such rigidity requirements, as illustrated in Fig.~\ref{fig4}. This motivates the investigation of formation control under sparse sensing conditions, where the available interactions may be insufficient to support conventional rigidity-based specifications.

Alternative geometric formation specifications offer different trade-offs under sparse sensing conditions, as summarized in Table~\ref{tab:comparison}. Displacement-based approaches can achieve formation control over a spanning-tree interaction graph with only $N-1$ edges, but they require a common orientation reference to specify the desired relative displacement vectors, either available a priori or estimated through additional orientation-alignment dynamics \cite{oh2014formation}. Bearing-based approaches preserve the formation scale as a free variable, but generally require bearing rigidity and a common orientation reference to determine the remaining geometric properties. In contrast, angle-based approaches do not require a common orientation reference and preserve global rotational freedom, but angle constraints alone cannot determine the formation scale. Existing distance--angle formulations further demonstrate that combining these constraints can determine both the shape and scale of a formation without requiring a rigid distance graph \cite{kwon2020generalized}. However, gradient feedback of subtended angles typically requires each agent to sense every vertex of the triangles in which it participates, involves configuration-dependent Jacobians that are undefined when the corresponding relative positions vanish, and its convergence guarantees are mostly local or limited to small formations \cite{kwon2020generalized,liu2020switching}. Table~\ref{tab:comparison} also includes these distance--angle methods. Compared with them, the proposed method needs only $N-1$ sensing links, at the price of exchanging scalars between neighbors, and its convergence is almost global for any $N$ without gain conditions, whereas the almost-global result of \cite{kwon2020generalized} is limited to three agents and \cite{liu2020switching} relies on state-dependent switching with shape-dependent gain conditions. These considerations motivate the exploration of an alternative formation-control architecture that exploits sparse relative measurements while preserving the desired formation scale and global rotational freedom.
\begin{table*}
\centering
\caption{Comparison of representative geometric formation specifications.}\label{tab:comparison}
\renewcommand{\arraystretch}{1.3}
\setlength{\tabcolsep}{3.5pt}
\begin{tabular}{>{\centering\arraybackslash}p{1.9cm} >{\centering\arraybackslash}p{1.9cm} >{\centering\arraybackslash}p{2.7cm} >{\centering\arraybackslash}p{2.3cm} >{\centering\arraybackslash}p{2.2cm} >{\centering\arraybackslash}p{3cm} >{\centering\arraybackslash}p{2.2cm}}
\hline\textbf{$\quad$}
& \textbf{Distance-based}\newline \cite{demarina2016distributed,krick2009stabilisation} & \textbf{Displacement-based}\newline \cite{de2020maneuvering,ren2007consensus} & \textbf{Bearing-based}\newline \cite{cheng2024general,su2026bearing} & \textbf{Angle-based}\newline \cite{chen2020anglesss,li2022satellite} & \textbf{Distance + angle}\newline \cite{kwon2020generalized,liu2020switching} & \textbf{Proposed method}\\
\hline\textbf{Formation specification} & $\Vert p_i^*-p_j^*\Vert$ &  $p_j^{*}-p_i^{*}$ & $\psi_{ij}^*=\dfrac{p_j^*-p_i^*}{\|p_j^*-p_i^*\|}$
&$\theta_{jik}^*=\angle(\psi_{ij}^*,\psi_{ik}^*)$
&$\Vert p_i^*-p_j^*\Vert$ and $\theta_{jik}^*$ (unsigned in \cite{kwon2020generalized}, signed in \cite{liu2020switching})
&Distances and relative virtual phases\\
\hline\textbf{Measurement}
& Distances
& Displacement in global frame
&$\psi_{ij}=\dfrac{p_j-p_i}{\|p_j-p_i\|}$
&$\{\psi_{ij},\psi_{ik}\}$, or $\theta_{jik}=\angle(\psi_{ij},\psi_{ik})$
&Displacement in local frame
&Displacement in local frame\\
\hline
\textbf{Global frame}& Not required& Required& Required & Not required; common handedness& Not required; common handedness in \cite{liu2020switching}& Not required; common handedness\\
\hline\textbf{Interaction graph}&Rigid ($\geq 2N-3$ edges)&Connected ($\geq N-1$ edges) &Bearing rigid ($\geq 2N-3$ edges)&Angle rigid ($\geq 2N-3$ edges)&Triangulated ($\geq 2N-3$ edges)&\textbf{Any tree ($N-1$ edges)}\\
\hline\textbf{Necessary communication}&None&None&None&None&None&One scalar per edge direction\\
\hline\textbf{Freedoms} &Translation and rotation &Translation &Translation and scale
&Translation, rotation and scale &Translation and rotation &Translation and rotation\\
\hline\textbf{Residual ambiguity}& Mirror-image ambiguity& None& Scale ambiguity & Scale ambiguity &Mirror image in \cite{kwon2020generalized}; none in \cite{liu2020switching} &None\\
\hline\textbf{Convergence guarantee}& Local; almost global for complete graphs \cite{park2018distance}& Global& Global (with leaders)& Local& \cite{kwon2020generalized}: local (almost global for $N=3$); \cite{liu2020switching}: almost global, with switching and a gain bound& \textbf{Almost global for any $N$; smooth, any gains}\\
\hline\end{tabular}
\end{table*}

Motivated by these considerations, we develop a formation-control framework inspired by parameterized Guiding Vector Fields (GVFs). In parameterized GVFs, a convergence component drives the physical state toward a prescribed manifold, while a virtual phase variable parameterizes its position along the manifold \cite{yao2021singularity}. This mechanism has been primarily exploited in path-following and path-coordination problems \cite{kapitanyuk2018guiding,yao2021singularity}. In this work, we extend the parameterization principle to formation control. Every internal agent of a tree interaction graph associates with each of its tree neighbors a circular manifold centered at itself. The convergence component regulates the radial deviation to establish the prescribed inter-agent distance, while the virtual phase describes the tangential degree of freedom along the circle. Coordinating the relative values of the phases hosted by the same agent establishes the internal formation geometry without imposing physical angle-gradient feedback or prescribing the global orientation.

The distributed implementation of the proposed framework introduces an additional challenge when agents operate in mutually misaligned local sensing frames. In particular, vector quantities expressed in one agent's local frame cannot be directly used by another agent without knowledge of their relative orientation. Existing approaches typically address this issue by estimating inter-frame orientations \cite{oh2014formation} or by constructing orientation-independent feedback quantities \cite{aranda2016distributed}. Following the latter principle, we develop a local-frame implementation in which every agent integrates the phases it hosts using only its own measurements, and one scalar per hosted phase is sent to the corresponding neighbor, which reconstructs its control input from its own locally measured relative displacement.

The main contributions of this work are summarized as follows:
\begin{itemize}
	\item A GVF-inspired formation-control framework is developed based on parameterized circular manifolds and virtual-phase coordination over an arbitrary tree interaction graph. The prescribed formation is achieved using exactly $N-1$ sensing links, preserving the desired formation scale and global rotational freedom. Since a tree is the sparsest connected graph, any connected set of available links suffices.
	\item The complete closed-loop equilibrium set is characterized. Undesired equilibria exist only in association with internal edges of the tree; they correspond to the collapse of a subset of such edges, and the Hessian of the associated potential has exactly as many negative eigenvalues as collapsed edges. Consequently, the desired formation is locally exponentially stable and almost globally attractive, and globally attractive when the tree is a star. An explicit energy level that excludes all undesired equilibria is also given.
	\item The controller is implemented under mutually misaligned local frames using locally expressed relative measurements and a single scalar per edge direction. The frames only need to share the same handedness, which is necessary for any method that, like ours, resolves the mirror-image ambiguity.
\end{itemize}

The remainder of this paper is organized as follows. Section~\ref{sec:prelim} gives the preliminaries, and Section~\ref{sec:problem} formulates the problem. Section~\ref{s3} develops the geometric representation and the graph condition, while Section~\ref{sec:control} presents the controller, the convergence analysis, and the local-frame implementation. Section~\ref{sec:sim} gives simulations, followed by conclusions in Section~\ref{sec:conclusion}.

\section{Preliminaries}\label{sec:prelim}

\subsection{Graph Theory}

Let $\mathcal G=(\mathcal V,\mathcal E)$ be an undirected graph with vertex set $\mathcal V=\{1,\ldots,N\}$ and edge set $\mathcal E\subseteq\{\{i,j\}:i,j\in\mathcal V,\,i\neq j\}$. The neighbor set of agent $i$ is denoted by $\mathcal N_i:=\{j\in\mathcal V:\{i,j\}\in\mathcal E\}$.
A graph is connected if there exists a path between every pair of vertices. 
A connected graph containing exactly $N-1$ edges is a tree; a tree whose vertex set is $\mathcal V$ and whose edges belong to $\mathcal E$ is a spanning tree of $\mathcal G$, and every connected graph contains one. 
In a tree, a vertex of degree one is a \emph{leaf}, and a vertex of degree at least two is \emph{internal}. 
A star is a tree with a single internal vertex.

For an oriented graph with $N$ vertices and $m$ edges, let $A\in\mathbb R^{m\times N}$ denote its incidence matrix, where each row corresponds to one oriented edge.
If the graph is a tree, then $m=N-1$, $A$ has full row rank and
$\ker(A)=\operatorname{span}\{\mathbf 1_N\}$.

\subsection{Planar Rotations}
Let $R(\theta):=
\begin{bmatrix}
	\cos\theta & -\sin\theta\\
	\sin\theta & \cos\theta
\end{bmatrix}
\in SO(2)$, $E:=R(\pi/2)$, and the unit vector $q(\theta):=\operatorname{col}(\cos\theta,\sin\theta)$, i.e., $\|q(\theta)\|=1$.
For planar rotations, one has
$R(\alpha)R(\beta)=R(\beta)R(\alpha)=R(\alpha+\beta)$, and rotations preserve Euclidean norms and inner products. Moreover,
\begin{equation}\label{eq:sin_identity}
	\big(Eq(\alpha)\big)^\top q(\beta)=\sin(\beta-\alpha),\qquad \alpha,\beta\in\mathbb R.
\end{equation}

For an arbitrary local frame $\mathcal O_i$, let
$F_i\in SO(2)$ denote its orientation relative to an arbitrary reference frame. Then the coordinate representation of a vector $v\in\mathbb R^2$
in $\mathcal O_i$ is
${}^iv=F_i^\top v$.

\section{Problem formulation}\label{sec:problem}

Consider a group of $N\geq 3$ agents evolving in the two-dimensional space, with the dynamics given by
\begin{equation}\label{eq:si}
	\dot p_i=u_i,\qquad i\in\mathcal V,
\end{equation}
where $p_i\in\mathbb R^2$ and $u_i\in\mathbb R^2$ denote the position and control input of agent $i$, respectively.
The interactions among the agents are described by a tree $\mathcal G=(\mathcal V,\mathcal E)$, $|\mathcal E|=N-1$, where $\{i,j\}\in\mathcal E$ indicates that agents $i$ and $j$ measure their relative displacement, each in its own local frame, and exchange scalar information.

Let $p^*=\operatorname{col}(p_1^*,\ldots,p_N^*)$ denote a prescribed formation configuration with $p_i^*\neq p_j^*$ whenever $\{i,j\}\in\mathcal E$. The objective is not to stabilize the agents to the particular configuration $p^*$, but to achieve the same formation shape while preserving the rigid-body translational and rotational freedoms. Accordingly, the desired formation set is defined as
\begin{equation}\label{stable_set}
	\mathcal S^*=\left\{\mathbf 1_N\otimes c+(I_N\otimes R(\alpha))p^*:
	c\in\mathbb R^2,\;\alpha\in\mathbb R\right\}.
\end{equation}
Hence, the control objective is to drive the multi-agent system toward $\mathcal S^*$ without prescribing either the absolute position or the absolute orientation of the formation.
\begin{problem}
	Consider the multi-agent system \eqref{eq:si} with a tree interaction graph $\mathcal G$ and a prescribed formation configuration $p^*$. Design distributed control laws that use only the locally expressed relative displacements over the $N-1$ edges of $\mathcal G$, together with scalar information exchanged along these edges, such that
	$$\operatorname{dist}(p(t),\mathcal S^*)\to0,
	\qquad t\to\infty,$$
	for almost all initial conditions.
\end{problem}
\begin{rem}
	The use of locally expressed relative displacement measurements for feedback should not be confused with displacement-based formation specifications. In the latter, desired relative displacement vectors are prescribed with respect to a common reference frame, whereas the proposed formulation uses local relative measurements only for controller implementation and does not prescribe an absolute formation orientation.
\end{rem}
\section{Geometric Representation and Graph Condition}\label{s3}
\subsection{Circular Parameterization and Relative Phases}
To characterize the geometric freedom remaining under a distance constraint, consider two neighboring agents $i$ and $j$. The constraint $\|p_j-p_i\|=d_{ij}^*$ confines agent $j$ to a circle of radius $d_{ij}^*$ centered at $p_i$, leaving one degree of freedom along the circle. Motivated by parameterized GVF methods, particularly the Singularity-Free GVF (SF-GVF) framework developed in \cite{yao2021singularity}, we parameterize this remaining degree of freedom by introducing a virtual phase. When agent $i$ hosts the circle associated with agent $j$, the corresponding desired position of agent $j$ is given by
\begin{equation}\label{eq:circle_parameterization}
	p_j^d(\omega_{j|i})=p_i+d_{ij}^*q(\omega_{j|i}),
\end{equation}
where $\omega_{j|i}\in\mathbb R$ denotes a lifted phase variable of agent $j$ with respect to the host agent $i$, and $q(\omega_{j|i})$ is the unit vector pointing from $p_i$ towards $p_j^d$. For brevity, we write $q_{j|i}:=q(\omega_{j|i})$.
\begin{figure}
	\centering
	\begin{tikzpicture}[scale=1.0, >=Stealth, every node/.style={font=\footnotesize},
		ag/.style={circle,draw,minimum size=5.5mm,inner sep=0pt}]
		\node[ag,fill=gray!25] (I) at (0,0) {$i$};
		\node[ag,fill=white] (J) at (35:2.75) {$j$};
		\coordinate (pd) at (75:2);
		\draw[gray!70,densely dotted,line width=0.8pt] (0,0) circle (2);
		\draw[dashed,gray!60!black] (I) -- (0:2.6);
		\node[below,align=center,font=\scriptsize,gray!50!black] at (1.6,0) {phase reference\\fixed in $\mathcal O_i$};
		\draw[->,thick] (I) -- node[left=-1pt,pos=0.6]{$d^*_{ij}q_{j|i}$} (pd);
		\fill (pd) circle (1.6pt) node[above left=-1pt]{$p^d_j$};
		\draw[->,thick,blue!70!black] (I) -- node[below right=-2pt,pos=0.55]{$r_{ij}$} (J);
		\draw[->,thick,red!75!black] (pd) -- node[above right=-2pt]{$\phi_{j|i}$} (J);
		\draw[->] (0:0.55) arc (0:75:0.55);
		\node at (14:0.95) {$\omega_{j|i}$};
		\draw[->,orange!85!black] (35:1.3) arc (35:75:1.3);
		\node[orange!85!black] at (57:1.6) {$\delta_{j|i}$};
		\draw[->,gray!70!black] (I) -- ++(130:0.75) node[left=-1pt,font=\scriptsize]{$x_i$};
		\draw[->,gray!70!black] (I) -- ++(220:0.75) node[left=-1pt,font=\scriptsize]{$y_i$};
		\draw[->,gray!70!black] (J) -- ++(45:0.75) node[above right=-2pt,font=\scriptsize]{$x_j$};
		\draw[->,gray!70!black] (J) -- ++(135:0.75) node[above left=-2pt,font=\scriptsize]{$y_j$};
		\draw[->,dashed,thick,orange!85!black] (I.south) .. controls (1.4,-1.5) and (3.9,-0.9) .. node[below right=-1pt]{sends $\delta_{j|i}$} (J.east);
	\end{tikzpicture}
	\caption{Information used by one arc $(i,j)$. The host $i$ measures $r_{ij}=p_j-p_i$ in its own frame $\mathcal O_i$ and integrates the virtual phase $\omega_{j|i}$, measured from a reference direction fixed in $\mathcal O_i$; the phase defines the point $p_j^d$ on the circle of radius $d_{ij}^*$, and $\phi_{j|i}=p_j-p_j^d$. The hosted agent $j$ measures the same displacement in its own frame $\mathcal O_j$, which need not be aligned with $\mathcal O_i$. The only transmitted quantity is the scalar $\delta_{j|i}$, the angle from $r_{ij}$ to $q_{j|i}$, which allows $j$ to express $\phi_{j|i}$ in $\mathcal O_j$.}
	\label{fig5}
\end{figure}
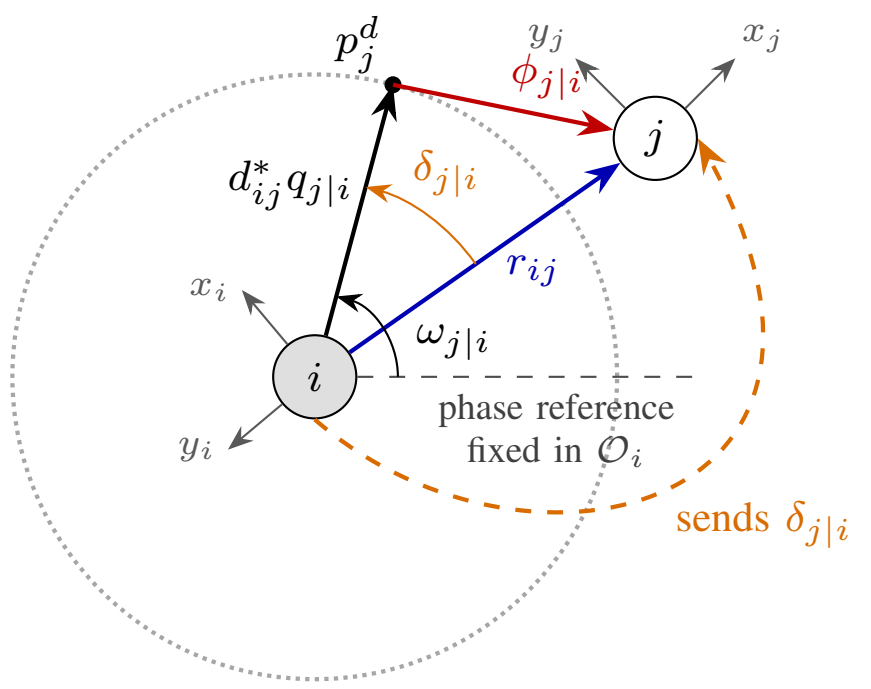
Accordingly, define the convergence error as $\phi_{j|i}=p_j-p_j^d$, then we have
\begin{equation}\label{eq:manifold_error}
	\phi_{j|i}=p_j-p_i-d_{ij}^*q_{j|i}.
\end{equation}
Thus, $\phi_{j|i}=0$ enforces the prescribed inter-agent distance while leaving the virtual phase $\omega_{j|i}$ free to parameterize the location of agent $j$ along the circular manifold. The reference direction corresponding to $\omega_{j|i}=0$ can be chosen arbitrarily by agent $i$ and need not be aligned with its local body frame, as illustrated in Fig.~\ref{fig5}.

To determine the relative geometry among multiple agents, the virtual phases associated with different circular manifolds cannot evolve independently. Phases hosted by the same agent share a common phase reference and can therefore be directly compared. Let agents $j$ and $k$ be associated with the same host agent $i$. Their relative-phase error is defined as
\begin{equation}\label{eq:relative_phase_error}
	e_{jk|i}=\omega_{j|i}-\omega_{k|i}-\Delta_{jk|i}^*,
\end{equation}
where $\Delta_{jk|i}^*$ denotes the desired signed phase difference determined by the prescribed formation. Specifically, let $\beta_{ij}^*\in\mathbb R$ satisfy $p_j^*-p_i^*=d_{ij}^*q(\beta_{ij}^*)$ and set $\Delta_{jk|i}^*:=\beta_{ij}^*-\beta_{ik}^*$; since the lifts $\beta^*_{ij}$ are defined up to multiples of $2\pi$, we choose $\Delta_{jk|i}^*\in(-\pi,\pi]$.
When the corresponding convergence errors $\phi_{j|i}$ and $\phi_{k|i}$ vanish, enforcing $e_{jk|i}=0$ fixes the signed angle subtended at agent $i$ by agents $j$ and $k$, while the common phase of the two remains unrestricted. Therefore, the prescribed distances determine the radial components of the formation, whereas the relative-phase relations coordinate the tangential components without prescribing the absolute orientation of the formation.

\subsection{Phase Hosting over a Spanning Tree}
\begin{figure}
	\centering
	\begin{tikzpicture}[>=Stealth, every node/.style={font=\footnotesize},
		ag/.style={circle,draw,minimum size=5.5mm,inner sep=0pt},
		host/.style={ag,fill=gray!25}]
		\node[ag]   (1) at (0,0)     {1};
		\node[host] (2) at (1.6,0.8) {2};
		\node[host] (3) at (3.2,0)   {3};
		\node[host] (4) at (4.8,0.8) {4};
		\node[ag]   (5) at (6.4,0)   {5};
		\node[ag]   (6) at (3.2,-1.5) {6};
		\draw[->,thick,red!70!black]   (2) -- node[above left=-1pt]{$\omega_{1|2}$} (1);
		\draw[->,thick,red!70!black]   (2) to[bend left=22] node[above right=-2pt]{$\omega_{3|2}$} (3);
		\draw[->,thick,blue!70!black]  (3) to[bend left=22] node[below left=-2pt]{$\omega_{2|3}$} (2);
		\draw[->,thick,blue!70!black]  (3) to[bend left=22] node[above left=-1pt]{$\omega_{4|3}$} (4);
		\draw[->,thick,blue!70!black]  (3) -- node[right]{$\omega_{6|3}$} (6);
		\draw[->,thick,green!45!black] (4) to[bend left=22] node[below right=-1pt]{$\omega_{3|4}$} (3);
		\draw[->,thick,green!45!black] (4) -- node[above right=-1pt]{$\omega_{5|4}$} (5);
	\end{tikzpicture}
	\caption{Phase hosting over a spanning tree. Internal agents (shaded) host one phase for each tree neighbor; arrows point from the host to the hosted agent. The internal edges $\{2,3\}$ and $\{3,4\}$ are parameterized from both endpoints, whereas the leaf edges are parameterized once, which gives $M=7$ phases. Relative-phase relations are imposed among the phases of each color, i.e., $1+2+1=N-2=4$ relations.}
	\label{fig:arcs}
\end{figure}
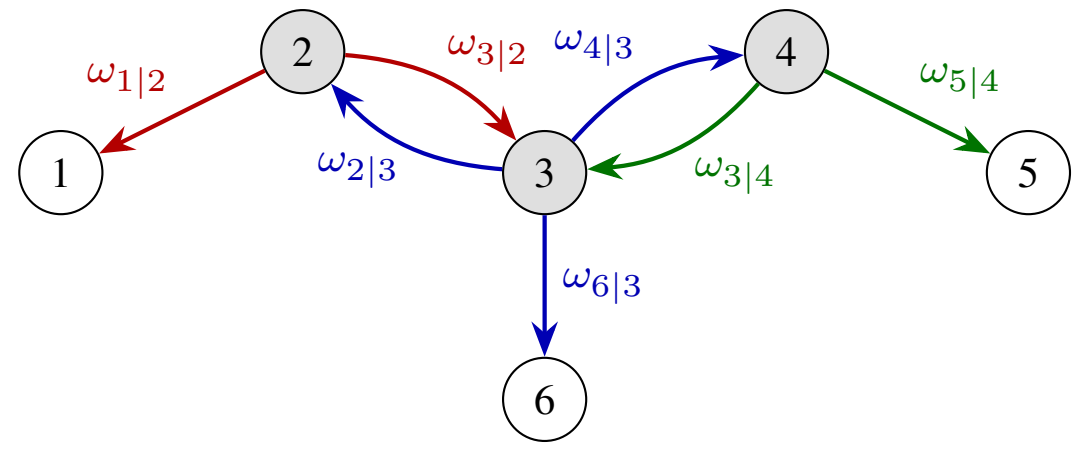

We now assign the virtual phases over the tree $\mathcal G$, where $\mathcal N_i$ denotes the neighbor set of agent $i$. Let $\mathcal V_I:=\{i\in\mathcal V:|\mathcal N_i|\geq 2\}$ and $\mathcal V_L:=\mathcal V\setminus\mathcal V_I$ denote the internal agents and the leaves of the tree, respectively. Every internal agent hosts one phase for each of its tree neighbors. We therefore define the set of arcs
\begin{equation}\label{eq:arcs}
	\mathcal A:=\big\{(i,j):i\in\mathcal V_I,\ j\in\mathcal N_i\big\},\qquad M:=|\mathcal A|,
\end{equation}
where, for $(i,j)\in\mathcal A$, agent $i$ is the host and agent $j$ the hosted agent of the phase $\omega_{j|i}$. The arcs hosted by agent $i$ are $\mathcal A_i:=\{(i,j):j\in\mathcal N_i\}$. Finally, the edges joining two internal agents are collected in
$$\mathcal E_I:=\big\{\{i,j\}\in\mathcal E:i,j\in\mathcal V_I\big\},$$
and we write $\mathcal G_I:=(\mathcal V_I,\mathcal E_I)$. An illustration is given in Fig.~\ref{fig:arcs}.

\begin{lem}\label{lem:counting}
	Let $\mathcal G$ be a tree with $N\geq3$ vertices. Then:
	\begin{enumerate}
		\item[(i)] $\mathcal V_I\neq\varnothing$, and $\mathcal G_I$ is a tree;
		\item[(ii)] every edge in $\mathcal E\setminus\mathcal E_I$ joins a leaf to an internal agent and induces exactly one arc, while every edge in $\mathcal E_I$ induces the two arcs $(i,j)$ and $(j,i)$; hence $M=2(N-1)-|\mathcal V_L|$;
		\item[(iii)] $\sum_{i\in\mathcal V_I}\big(|\mathcal N_i|-1\big)=N-2$.
	\end{enumerate}
\end{lem}
\begin{proof}
	(i) If no vertex had degree at least two, every component of $\mathcal G$ would contain at most two vertices, contradicting connectivity for $N\geq 3$. The interior vertices of any path have degree at least two; hence the path of $\mathcal G$ joining two internal agents only visits internal agents, and $\mathcal G_I$ is connected. It is acyclic as a subgraph of a tree. 
	(ii) An edge joining two leaves would form a connected component with two vertices, which is impossible for $N\geq3$. The arc count follows from $M=\sum_{i\in\mathcal V_I}|\mathcal N_i|=\sum_{i\in\mathcal V}|\mathcal N_i|-|\mathcal V_L|=2(N-1)-|\mathcal V_L|$. 
	(iii) $\sum_{i\in\mathcal V_I}(|\mathcal N_i|-1)=M-|\mathcal V_I|=2N-2-|\mathcal V_L|-(N-|\mathcal V_L|)=N-2$.
\end{proof}

For each host $i\in\mathcal V_I$, the phases $\{\omega_{j|i}\}_{(i,j)\in\mathcal A_i}$ share the same reference. To obtain an independent set of relations without redundancy, we select a spanning tree over these $|\mathcal N_i|$ phases (for instance, a path) and impose the corresponding $|\mathcal N_i|-1$ relations \eqref{eq:relative_phase_error}. By Lemma~\ref{lem:counting}(iii), this yields exactly $N-2$ relations. Collecting the phases as $\omega=\operatorname{col}(\omega_{j|i})_{(i,j)\in\mathcal A}\in\mathbb R^M$, the relations are written compactly as
\begin{equation}\label{eq:phase_error_compact}
	e_\omega=A_\omega\omega-\Delta^*,
\end{equation}
where $\Delta^*\in\mathbb R^{N-2}$ collects the desired phase differences and $A_\omega\in\mathbb R^{(N-2)\times M}$ is block diagonal, its $i$-th block being the incidence matrix of the tree selected over $\mathcal A_i$. Hence $A_\omega$ has full row rank, and
\begin{equation}\label{eq:kerA}
	\ker(A_\omega)=\operatorname{span}\{\mathbf 1_{\mathcal A_i}:i\in\mathcal V_I\},
\end{equation}
where $\mathbf 1_{\mathcal A_i}\in\mathbb R^M$ is the indicator vector of the arcs hosted by agent $i$. Together with the $N-1$ distances, the relations provide $2N-3$ scalar constraints, which matches the $2N-3$ degrees of freedom of a planar configuration modulo translations and rotations. The next result shows that they determine the formation.

\begin{proposition}\label{prop:shape}
	Let $\mathcal G$ be a tree with $N\geq 3$ vertices. If $\phi_{j|i}=0$ for all $(i,j)\in\mathcal A$ and $e_\omega=0$, then $p\in\mathcal S^*$. Conversely, for every $p\in\mathcal S^*$ there exists $\omega\in\mathbb R^M$ such that $\phi_{j|i}=0$ for all $(i,j)\in\mathcal A$ and $e_\omega=0$.
\end{proposition}
\begin{proof}
	Since the relations selected over $\mathcal A_i$ form a tree, $e_\omega=0$ implies the existence of $\theta_i\in\mathbb R$ such that $\omega_{j|i}=\theta_i+\beta_{ij}^*$ for all $(i,j)\in\mathcal A_i$. Then, $\phi_{j|i}=0$ gives
	\begin{equation}\label{eq:arc_rotation}
		p_j-p_i=d_{ij}^*q(\theta_i+\beta_{ij}^*)=R(\theta_i)(p_j^*-p_i^*),\qquad (i,j)\in\mathcal A.
	\end{equation}
	For $\{i,j\}\in\mathcal E_I$, both arcs $(i,j)$ and $(j,i)$ exist, and \eqref{eq:arc_rotation} yields $R(\theta_i)(p_j^*-p_i^*)=-R(\theta_j)(p_i^*-p_j^*)$. Since $p_j^*\neq p_i^*$, it follows that $R(\theta_i)=R(\theta_j)$. Because $\mathcal G_I$ is connected by Lemma~\ref{lem:counting}(i), $R(\theta_i)=R(\alpha)$ for some $\alpha\in\mathbb R$ and all $i\in\mathcal V_I$. Every edge of $\mathcal E$ induces at least one arc; hence $p_j-p_i=R(\alpha)(p_j^*-p_i^*)$ for all $\{i,j\}\in\mathcal E$. Since $\mathcal G$ is connected, $p=\mathbf 1_N\otimes c+(I_N\otimes R(\alpha))p^*$ with $c=p_1-R(\alpha)p_1^*$, i.e., $p\in\mathcal S^*$. Conversely, for $p=\mathbf 1_N\otimes c+(I_N\otimes R(\alpha))p^*$, the choice $\omega_{j|i}=\alpha+\beta_{ij}^*$ satisfies $\phi_{j|i}=0$ and $e_\omega=0$.
\end{proof}

\begin{rem}[Choice of the tree]\label{rem:tree_choice}
	If more links than a tree are available, any spanning tree of them can be used (Corollary~\ref{cor:connected}). As shown in Section~\ref{sec:control}, undesired equilibria are associated only with the $|\mathcal E_I|=|\mathcal V_I|-1=N-1-|\mathcal V_L|$ internal edges. Hence, trees with many leaves are preferable from the viewpoint of the equilibrium structure; in the extreme case of a star, $\mathcal E_I=\varnothing$. Conversely, a star requires a single agent to sense all others, which is precisely what LoS restrictions may prevent. If, instead, every agent were assigned exactly one host, as in single-center constructions, the requirement of $N-2$ independent relations would force a double-star structure, i.e., two adjacent agents sensing all the others; letting each internal agent host one phase per tree neighbor removes this restriction.
\end{rem}

\section{Controller Design and Convergence Analysis}\label{sec:control}

\subsection{GVF-Based Formation Controller}\label{sectionIV-A}
Following the augmented-space construction of SF-GVF \cite{yao2021singularity}, we first formulate a convergence vector field associated with each parametric constraint \eqref{eq:manifold_error}. For an arc $(i,j)\in\mathcal A$, consider the augmented coordinate $\xi_{j|i}=\operatorname{col}\big(p_j,p_i,\omega_{j|i}\big)\in\mathbb R^5$.
The Jacobian of $\phi_{j|i}$ with respect to $\xi_{j|i}$ is
\begin{equation}\label{eq:local_jacobian}
	J_{j|i}:=\frac{\partial\phi_{j|i}}{\partial\xi_{j|i}}=
	\begin{bmatrix}
		I_2 & -I_2 & -\tau_{j|i}
	\end{bmatrix},\quad \tau_{j|i}:=d_{ij}^*Eq_{j|i},
\end{equation}
and, following \cite{yao2021singularity}, we define the convergence field associated with $\phi_{j|i}$ as
\begin{equation}\label{eq:local_convergence_field}
	\chi_{j|i}^{\rm con}=-k_pJ_{j|i}^{\top}\phi_{j|i}=\begin{bmatrix}
		-k_p\phi_{j|i}\\
		k_p\phi_{j|i}\\
		k_p\tau_{j|i}^\top\phi_{j|i}
	\end{bmatrix},
\end{equation}
where the constant $k_p>0$.
Hence, the parameterized circular path associated with the arc $(i,j)$
contributes $-k_p\phi_{j|i}$ to the motion of the hosted agent $j$, the opposite term
$k_p\phi_{j|i}$ to its host $i$, and
$k_p\tau_{j|i}^\top\phi_{j|i}$ to the evolution of the corresponding virtual
phase.

In conventional SF-GVF constructions, an additional tangential component may be introduced to generate persistent motion along the prescribed path. Such a tangential component is not required here, since the objective is to stabilize a static formation rather than to drive the agents continuously along their circular reference paths. We therefore set the tangential GVF component to $\chi_{j|i}^{\rm tan}=0\in\operatorname{Ker}(J_{j|i})$, and exploit the additional degree of freedom provided by the virtual phase coordinate to enforce the relative-phase relations introduced in the previous section. To this end, we add the virtual-phase term $\chi_{j|i}^\omega=\operatorname{col}\big(0_4,-k_\omega[A_\omega^\top e_\omega]_{(i,j)}\big)$ with $k_\omega>0$, where $[\,\cdot\,]_{(i,j)}$ denotes the entry associated with the arc $(i,j)$.

For the single-integrator dynamics considered in this paper, the physical components of the constructed vector fields can be directly realized as the control inputs. By superimposing all field components acting on agent $k$, the controlled system is given by
\begin{subequations}\label{eq:proposed_controller}
	\begin{align}
		\dot p_k&=u_k
		=-k_p\!\!\sum_{i:(i,k)\in\mathcal A}\!\!\phi_{k|i}+k_p\!\!\sum_{j:(k,j)\in\mathcal A}\!\!\phi_{j|k} ,
		\label{eq:proposed_position_controller}\\
		\dot\omega_{j|i}&=k_p\tau_{j|i}^\top\phi_{j|i}-
		k_\omega[A_\omega^\top e_\omega]_{(i,j)},\quad (i,j)\in\mathcal A.
		\label{eq:proposed_phase_controller}
	\end{align}
\end{subequations}
For a leaf $k$, the first sum in \eqref{eq:proposed_position_controller} contains the single arc hosted by its unique neighbor and the second sum is empty; for an internal agent, the first sum runs over its internal neighbors and the second over all its tree neighbors. The physical dynamics are generated solely by the convergence fields and vanish once the corresponding manifold errors become zero; hence they do not generate persistent motion along the circles. The virtual-phase dynamics adjust the parameterized points along the circular manifolds, and the relative-phase coordination term provides the tangential adjustment required to establish the prescribed phase relations.

\subsection{Convergence Analysis}
To analyze the closed-loop behavior \eqref{eq:proposed_controller}, we define the potential function as
\begin{equation}\label{eq:potential_function}
	\mathcal H(p,\omega)=\frac{k_p}{2}\sum_{(i,j)\in\mathcal A}\|\phi_{j|i}\|^2
	+\frac{k_\omega}{2}\|e_\omega\|^2 .
\end{equation}
Let $x:=\operatorname{col}(p,\omega)\in\mathbb R^{2N+M}$ denote the augmented state, and define the equilibrium set of the closed-loop system as
$\mathcal Z:=\left\{x:\dot p=0,\;\dot\omega=0\right\}$. Since each $\phi_{j|i}$ is $2\pi$-periodic in $\omega_{j|i}$ and $A_\omega\mathbf 1_{\mathcal A_i}=0$, both $\mathcal H$ and the closed loop are invariant under the shifts $\omega\mapsto\omega+2\pi\mathbf 1_{\mathcal A_i}$, $i\in\mathcal V_I$. We denote by $\Lambda:=\{2\pi\sum_{i\in\mathcal V_I}n_i\mathbf 1_{\mathcal A_i}:n_i\in\mathbb Z\}$ the lattice generated by these shifts.

\begin{thm}\label{thm:equilibrium_convergence}
	Consider the closed-loop system \eqref{eq:proposed_controller}. Then $\dot x=-\nabla\mathcal H(x)$, and 
	\begin{equation}\label{eq:Hdot}
		\dot{\mathcal H}
		=
		-\|\dot p\|^2
		-\|\dot\omega\|^2
		\leq 0,
	\end{equation}
	and the centroid $\frac1N\sum_{k=1}^Np_k$ is invariant. Moreover, every closed-loop trajectory exists for all $t\geq0$, $p(t)$ is bounded, and $x(t)$ converges to a single point of the equilibrium set $\mathcal Z$.
\end{thm}

\begin{proof}
	From \eqref{eq:manifold_error}, $\phi_{k|i}$ depends on $p_k$ with Jacobian $I_2$ and $\phi_{j|k}$ depends on $p_k$ with Jacobian $-I_2$; hence we have $\nabla_{p_k}\mathcal H=k_p\sum_{i:(i,k)\in\mathcal A}\phi_{k|i}-k_p\sum_{j:(k,j)\in\mathcal A}\phi_{j|k}=-\dot p_k$. Similarly, $\partial\phi_{j|i}/\partial\omega_{j|i}=-\tau_{j|i}$ gives $\nabla_{\omega_{j|i}}\mathcal H=-k_p\tau_{j|i}^\top\phi_{j|i}+k_\omega[A_\omega^\top e_\omega]_{(i,j)}=-\dot\omega_{j|i}$. Thus, we obtain $\dot x=-\nabla\mathcal H(x)$, which implies \eqref{eq:Hdot}. Each arc contributes $-k_p\phi_{j|i}$ to $\dot p_j$ and $+k_p\phi_{j|i}$ to $\dot p_i$, so $\sum_{k=1}^N\dot p_k=0$.

	Since $\mathcal H(x(t))\leq\mathcal H(x(0))$, we have $\|\phi_{j|i}\|\leq\sqrt{2\mathcal H(x(0))/k_p}$, and hence $\|p_j-p_i\|\leq d_{ij}^*+\sqrt{2\mathcal H(x(0))/k_p}$ for every $\{i,j\}\in\mathcal E$, because every tree edge induces at least one arc. As $\mathcal G$ is connected and the centroid is invariant, $p(t)$ is bounded. Decompose $\omega=\omega^\perp+\sum_{i\in\mathcal V_I}n_i\mathbf 1_{\mathcal A_i}$ with $\omega^\perp\perp\ker(A_\omega)$. Since $A_\omega\omega^\perp=e_\omega+\Delta^*$ is bounded and $A_\omega$ is injective on $\ker(A_\omega)^\perp$, $\omega^\perp$ is bounded. Hence, $\dot\omega$ is bounded, the solution exists for all $t\geq 0$, and the trajectory is precompact in the quotient $\mathcal X:=\mathbb R^{2N}\times(\mathbb R^M/\Lambda)$, on which the closed loop is the gradient flow of the real-analytic function induced by $\mathcal H$. By the {\L}ojasiewicz gradient inequality \cite{lojasiewicz1963propriete,absil2005convergence}, the projected trajectory converges to a single point of $\mathcal X$, which is necessarily an equilibrium. As $\Lambda$ is discrete and $x(t)$ is continuous, $x(t)$ itself converges to a single point of $\mathcal Z$.
\end{proof}

Next, we analyze the equilibria. The following elementary property of trees is used repeatedly.

\begin{lem}\label{lem:tree_flow}
	Let $\mathcal T=(\mathcal V_T,\mathcal E_T)$ be a tree, and let $f_{ij}\in\mathbb R^m$ be defined for every ordered pair of adjacent vertices, with $f_{ij}=-f_{ji}$. If $\sum_{j:\{i,j\}\in\mathcal E_T}f_{ij}=0$ for all $i\in\mathcal V_T$, then $f_{ij}=0$ for all $\{i,j\}\in\mathcal E_T$.
\end{lem}
\begin{proof}
	The claim is trivial for a single vertex. Otherwise, a leaf $i$ of $\mathcal T$ has a single neighbor $j$, so $f_{ij}=0$. Removing $i$ yields a tree on which the hypothesis still holds, since the removed term vanishes; the result follows by induction on $|\mathcal V_T|$.
\end{proof}

Recall that $x=\operatorname{col}(p,\omega)\in\mathbb R^{2N+M}$ is the augmented state, which stacks the positions $p$ of all agents and the phases $\omega$ hosted by all internal agents; the quantities $e_\omega$, $\phi_{j|i}$, and $q_{j|i}$ below are evaluated at $x$ through \eqref{eq:manifold_error} and \eqref{eq:phase_error_compact}. For every subset $\mathcal C\subseteq\mathcal E_I$ of internal edges, define
\begin{equation}\label{eq:E_S}
	\mathcal Z_{\mathcal C}:=\left\{x=\operatorname{col}(p,\omega):
	\begin{array}{l}
		e_\omega=0,\\[0.5mm]
		\phi_{j|i}=0,\ \ (i,j)\in\mathcal A,\ \{i,j\}\notin\mathcal C,\\[0.5mm]
		p_i=p_j,\ q_{j|i}=q_{i|j},\ \ \{i,j\}\in\mathcal C
	\end{array}
	\right\}.
\end{equation}
In words, a state belongs to $\mathcal Z_{\mathcal C}$ when all relative-phase errors vanish, every arc outside $\mathcal C$ lies on its circle, and the two endpoints of every edge in $\mathcal C$ coincide while their two phases point in the same direction. On such a collapsed edge $\{i,j\}\in\mathcal C$, \eqref{eq:manifold_error} gives $\phi_{j|i}=\phi_{i|j}=-d_{ij}^*q_{j|i}$. The desired equilibrium set is $\mathcal Z_d:=\mathcal Z_{\varnothing}=\{x:\phi_{j|i}=0\ \forall(i,j)\in\mathcal A,\ e_\omega=0\}$, whose projection onto the positions is $\mathcal S^*$ by Proposition~\ref{prop:shape}.

\begin{thm}\label{thm:equilibrium_characterization}
	The equilibrium set of the closed-loop system \eqref{eq:proposed_controller} is the disjoint union
	\begin{equation}\label{eq:E_decomposition}
		\mathcal Z=\bigcup_{\mathcal C\subseteq\mathcal E_I}\mathcal Z_{\mathcal C}.
	\end{equation}
	In particular, every undesired equilibrium corresponds to the collapse of a nonempty set of internal edges, and $e_\omega=0$ at every equilibrium.
\end{thm}

\begin{proof}
	Let $x\in\mathcal Z$. \emph{Leaves.} For a leaf $\ell$ with neighbor $h$, \eqref{eq:proposed_position_controller} reads $\dot p_\ell=-k_p\phi_{\ell|h}$, hence $\phi_{\ell|h}=0$.

	\emph{Internal edges: positions.} For $k\in\mathcal V_I$, the arcs hosted by $k$ towards leaves vanish, and $\dot p_k=0$ reduces to
	$\sum_{j\in\mathcal N_k\cap\mathcal V_I}\big(\phi_{j|k}-\phi_{k|j}\big)=0$. The quantities $f_{kj}:=\phi_{j|k}-\phi_{k|j}$ are antisymmetric on the tree $\mathcal G_I$, so Lemma~\ref{lem:tree_flow} gives $\phi_{j|k}=\phi_{k|j}=:z_{kj}$ for every $\{k,j\}\in\mathcal E_I$. Using \eqref{eq:manifold_error} for both arcs, this is equivalent to
	\begin{align}
		p_j-p_k&=\frac{d_{kj}^*}{2}(q_{j|k}-q_{k|j}),
		\label{eq:center_relative_equilibrium}\\
		z_{kj}&=-\frac{d_{kj}^*}{2}(q_{j|k}+q_{k|j}).
		\label{eq:center_z_equilibrium}
	\end{align}

	\emph{Internal edges: phases.} Let $g\in\mathbb R^M$ with $g_{(i,j)}:=\tau_{j|i}^\top\phi_{j|i}$. The condition $\dot\omega=0$ reads $k_pg=k_\omega A_\omega^\top e_\omega$. Premultiplying by $\mathbf 1_{\mathcal A_k}^\top$ and using \eqref{eq:kerA} gives $\sum_{j\in\mathcal N_k}g_{(k,j)}=0$ for every $k\in\mathcal V_I$, where the entries associated with leaves vanish. For $\{k,j\}\in\mathcal E_I$, \eqref{eq:sin_identity} and \eqref{eq:center_z_equilibrium}  give
	\begin{equation}\label{eq:g_antisym}
		g_{(k,j)}=-\frac{(d_{kj}^*)^2}{2}\sin(\omega_{k|j}-\omega_{j|k})=-g_{(j,k)}.
	\end{equation}
	Applying Lemma~\ref{lem:tree_flow} to $\mathcal G_I$ again yields $g=0$ and $\sin(\omega_{k|j}-\omega_{j|k})=0$ for every internal edge. Consequently, $A_\omega^\top e_\omega=0$ and, since $A_\omega$ has full row rank, $e_\omega=0$.

	\emph{Classification.} For every $\{k,j\}\in\mathcal E_I$, either $q_{k|j}=-q_{j|k}$, and then $z_{kj}=0$ by \eqref{eq:center_z_equilibrium}; or $q_{k|j}=q_{j|k}$, and then $p_j=p_k$ by \eqref{eq:center_relative_equilibrium}. Letting $\mathcal C$ collect the internal edges of the second type, $x\in\mathcal Z_{\mathcal C}$. Conversely, at any $x\in\mathcal Z_{\mathcal C}$, the arcs towards leaves vanish, $\phi_{j|k}=\phi_{k|j}$ on every internal edge, $g=0$ because $\tau_{j|i}^\top q_{j|i}=0$, and $e_\omega=0$; hence $\dot p=0$ and $\dot\omega=0$. The union of the sets $\mathcal Z_{\mathcal C}$ is disjoint because $p_i=p_j$ on $\{i,j\}\in\mathcal C$, whereas $\|p_i-p_j\|=d_{ij}^*>0$ on internal edges outside $\mathcal C$.
\end{proof}

For the local analysis of the equilibria, note that $\partial^2\phi_{j|i}/\partial\omega_{j|i}^2=d_{ij}^*q_{j|i}$. Hence, for perturbations $v=\operatorname{col}(v_k)\in\mathbb{R}^{2N}$ of the positions and $a=\operatorname{col}(a_{j|i})\in\mathbb{R}^{M}$ of the phases, the Hessian of $\mathcal H$ at an equilibrium $x^*=\operatorname{col}(p^\circ,\omega^\circ)$ defines the quadratic form
\begin{align}
	Q(v,a)=\,&k_p\sum_{(i,j)\in\mathcal A}\big\|v_j-v_i-\tau_{j|i}^*a_{j|i}\big\|^2
	+k_\omega\|A_\omega a\|^2\nonumber\\
	&+k_p\sum_{(i,j)\in\mathcal A}d_{ij}^*\big(q_{j|i}^{*\top}\phi_{j|i}^*\big)a_{j|i}^2.
	\label{eq:hessian_quadratic_form}
\end{align}
where the superscript $*$ on $\tau$, $q$, and $\phi$ denotes evaluation at $x^*$. At $x^*\in\mathcal Z_{\mathcal C}$, the last sum equals $-k_p\sum_{\{i,j\}\in\mathcal C}(d_{ij}^*)^2(a_{j|i}^2+a_{i|j}^2)$, since it only involves the collapsed arcs.

\begin{proposition}\label{prop:local_exp}
	At every $x^*\in\mathcal Z_d$, $\nabla^2\mathcal H(x^*)$ is positive semidefinite and its kernel is the three-dimensional tangent space of $\mathcal Z_d$,
	\begin{equation*}
		\ker\nabla^2\mathcal H(x^*)
		=\left\{\begin{bmatrix}\mathbf 1_N\otimes \eta+\gamma (I_N\otimes E)p^{\circ}\\[1mm]
			\gamma\mathbf 1_M
		\end{bmatrix}:\eta\in\mathbb R^2,\ \gamma\in\mathbb R
		\right\},
	\end{equation*}
	where $p^\circ$ denotes the position component of $x^*$. Consequently, $\mathcal Z_d$ is a normally hyperbolic attracting manifold of equilibria and is locally exponentially stable.
\end{proposition}
\begin{proof}
	At $x^*\in\mathcal Z_d$, all $\phi^*_{j|i}$ vanish, so $Q\geq0$, and $Q(v,a)=0$ if and only if $A_\omega a=0$ and $v_j-v_i=\tau_{j|i}^*a_{j|i}$ for all arcs. 
    By \eqref{eq:kerA}, $a_{j|i}=\gamma_i$ for all $(i,j)\in\mathcal A_i$. Since $p^\circ_j-p^\circ_i=d_{ij}^*q_{j|i}^*$, we have $\tau_{j|i}^*=E(p^\circ_j-p^\circ_i)$. 
    For $\{i,j\}\in\mathcal E_I$, both arcs give $v_j-v_i=\gamma_iE(p^\circ_j-p^\circ_i)$ and $v_i-v_j=\gamma_jE(p^\circ_i-p^\circ_j)$, so that $\gamma_i=\gamma_j$. As $\mathcal G_I$ is connected, $\gamma_i=\gamma$ for all hosts, and $v_j-v_i=\gamma E(p^\circ_j-p^\circ_i)$ on every tree edge, which yields the stated kernel. By Proposition~\ref{prop:shape}, $\mathcal Z_d$ consists of the points with $p=\mathbf 1_N\otimes c+(I_N\otimes R(\alpha))p^*$ and $\omega_{j|i}=\alpha+\beta^*_{ij}+2\pi n_i$, with $c\in\mathbb R^2$, $\alpha\in\mathbb R$, and $n_i\in\mathbb Z$ for each host $i$; hence it is a three-dimensional manifold whose tangent space coincides with the above kernel. Since the linearization of $\dot x=-\nabla\mathcal H(x)$ at $x^*$ is $-\nabla^2\mathcal H(x^*)$, the transverse spectrum is strictly negative, and it is the same at all points of a connected component of $\mathcal Z_d$ by the rigid-motion invariance of the closed loop.
\end{proof}

\begin{thm}\label{thm:saddle}
	Let $\mathcal C\subseteq\mathcal E_I$ and $x^*=\operatorname{col}(p^\circ,\omega^\circ)\in\mathcal Z_{\mathcal C}$. Then $\nabla^2\mathcal H(x^*)$ has exactly $|\mathcal C|$ negative eigenvalues, three zero eigenvalues, and $2N+M-|\mathcal C|-3$ positive eigenvalues. Moreover, $\ker\nabla^2\mathcal H(x^*)$ is spanned by the two translations and the common rotation, i.e., it is given by the expression in Proposition~\ref{prop:local_exp}. Consequently, for $\mathcal C\neq\varnothing$, $\mathcal Z_{\mathcal C}$ is a normally hyperbolic saddle manifold of equilibria with $|\mathcal C|$ unstable directions.
\end{thm}
\begin{proof}
	\emph{At most $|\mathcal C|$.} Consider $\{i,j\}\in\mathcal C$ and let $w:=v_j-v_i$, $\tau^*:=\tau_{j|i}^*=\tau_{i|j}^*$, where the equality follows from $q^*_{j|i}=q^*_{i|j}$. Using $\|\tau^*\|=d_{ij}^*$, the terms of \eqref{eq:hessian_quadratic_form} associated with the two collapsed arcs can be rewritten exactly as
	\begin{align*}
		&\big\|w-\tau^*a_{j|i}\big\|^2+\big\|w+\tau^*a_{i|j}\big\|^2-(d_{ij}^*)^2(a_{j|i}^2+a_{i|j}^2)\\
		&=2\Big\|w-\frac{\tau^*}{2}(a_{j|i}-a_{i|j})\Big\|^2-\frac{(d_{ij}^*)^2}{2}(a_{j|i}-a_{i|j})^2.
	\end{align*}
	Therefore, $Q=Q_+-\frac{k_p}{2}\sum_{\{i,j\}\in\mathcal C}(d_{ij}^*)^2(a_{j|i}-a_{i|j})^2$, where $Q_+\geq0$. 
    If $Q$ were negative definite on a subspace of dimension $|\mathcal C|+1$, this subspace would contain a nonzero vector annihilating the $|\mathcal C|$ linear functionals $a_{j|i}-a_{i|j}$, on which $Q=Q_+\geq0$, a contradiction.

	\emph{At least $|\mathcal C|$.} Removing the edges of $\mathcal C$ from $\mathcal G$ yields $|\mathcal C|+1$ subtrees with vertex sets $\mathcal V^{(0)},\ldots,\mathcal V^{(|\mathcal C|)}$, and contracting each of them yields a tree $\mathcal T_{\mathcal C}$ whose edges are the elements of $\mathcal C$. Each $\mathcal V^{(\ell)}$ contains an endpoint of a collapsed edge, which is an internal agent. 
    Given $\rho\in\mathbb R^{|\mathcal C|+1}$, perturb each subtree by an infinitesimal rigid motion, $v_k=\rho_\ell Ep^\circ_k+t_\ell$ for $k\in\mathcal V^{(\ell)}$, and $a_{j|i}=\rho_\ell$ for every arc hosted by $i\in\mathcal V^{(\ell)}$. On every arc not associated with $\mathcal C$, $v_j-v_i-\tau_{j|i}^*a_{j|i}=\rho_\ell E\phi^*_{j|i}=0$, and $A_\omega a=0$ because $a$ is constant over each $\mathcal A_i$. 
    For $\{i,j\}\in\mathcal C$ with $i\in\mathcal V^{(\ell)}$ and $j\in\mathcal V^{(m)}$, the translations can be assigned recursively along $\mathcal T_{\mathcal C}$, linearly in $\rho$, so that $w=\frac{\tau^*}{2}(\rho_\ell-\rho_m)$. Then $Q_+=0$ and
	$$Q=-\frac{k_p}{2}\sum_{\{i,j\}\in\mathcal C}(d_{ij}^*)^2(\rho_\ell-\rho_m)^2,$$
	which is negative definite on the $|\mathcal C|$-dimensional subspace $\{\rho:\mathbf 1^\top\rho=0\}$, because the incidence matrix of the tree $\mathcal T_{\mathcal C}$ has rank $|\mathcal C|$ and its kernel is spanned by $\mathbf 1$. The map $\rho\mapsto(v,a)$ is injective, since every $\mathcal V^{(\ell)}$ contains a host. Hence $\nabla^2\mathcal H(x^*)$ has at least $|\mathcal C|$ negative eigenvalues.

	\emph{Kernel.} Since $\mathcal H$ is invariant under common translations and under rotating all positions while shifting all phases by the same angle, the three corresponding generators belong to $\ker\nabla^2\mathcal H(x^*)$. 
    Conversely, let $(v,a)\in\ker\nabla^2\mathcal H(x^*)$. 
    This is the linearization of the equilibrium conditions at $x^*$, so we repeat the steps of the proof of Theorem~\ref{thm:equilibrium_characterization} for $\delta\phi_{j|i}:=v_j-v_i-\tau_{j|i}^*a_{j|i}$ and for the linearization $\delta g$ of $g$. Leaves give $\delta\phi_{\ell|h}=0$, and Lemma~\ref{lem:tree_flow} gives $\delta\phi_{j|k}=\delta\phi_{k|j}$ on every internal edge. Using $\partial\tau_{j|i}/\partial\omega_{j|i}=-d_{ij}^*q_{j|i}$, we have $\delta g_{(i,j)}=\tau_{j|i}^{*\top}\delta\phi_{j|i}$ on non-collapsed arcs and $\delta g_{(i,j)}=(d_{ij}^*)^2a_{j|i}+\tau_{j|i}^{*\top}\delta\phi_{j|i}$ on collapsed arcs. On $\{k,j\}\in\mathcal C$, the equality $\delta\phi_{j|k}=\delta\phi_{k|j}$ gives $w=\frac{\tau^*}{2}(a_{j|k}-a_{k|j})$ and $\delta g_{(k,j)}=\frac{(d_{kj}^*)^2}{2}(a_{j|k}-a_{k|j})=-\delta g_{(j,k)}$; on $\{k,j\}\in\mathcal E_I\setminus\mathcal C$, $\tau^*_{k|j}=-\tau^*_{j|k}$ gives $\delta g_{(j,k)}=-\delta g_{(k,j)}$. The linearized phase equation $k_p\delta g=k_\omega A_\omega^\top A_\omega a$, premultiplied by $\mathbf 1_{\mathcal A_k}^\top$, yields $\sum_{j}\delta g_{(k,j)}=0$, so Lemma~\ref{lem:tree_flow} gives $\delta g=0$. Then $A_\omega a=0$, i.e., $a_{j|i}=\gamma_i$ on $\mathcal A_i$, and $\delta g=0$ on a collapsed edge gives $\gamma_k=\gamma_j$. On a non-collapsed internal edge, $\delta\phi_{j|k}=\delta\phi_{k|j}$ gives $\delta\phi_{j|k}=\frac{\tau^*_{j|k}}{2}(\gamma_j-\gamma_k)$, and $\delta g_{(k,j)}=0$ gives $\gamma_j=\gamma_k$. As $\mathcal G_I$ is connected, $\gamma_i=\gamma$ for all hosts. 
    Consequently, $\delta\phi_{j|i}=0$ on all non-collapsed arcs and $w=0$ on collapsed edges, i.e., $v_j-v_i=\gamma E(p^\circ_j-p^\circ_i)$ on every edge of $\mathcal E$, because $\tau^*_{j|i}=E(p^\circ_j-p^\circ_i)$ on non-collapsed arcs and $p^\circ_i=p^\circ_j$ on collapsed ones. Hence $v=\mathbf 1_N\otimes\eta+\gamma(I_N\otimes E)p^\circ$ and $a=\gamma\mathbf 1_M$, and the kernel is three-dimensional. Together with the count of negative eigenvalues, this proves the claimed numbers of negative, zero, and positive eigenvalues.

	Finally, the orbit of $x^*$ under rigid motions is a three-dimensional manifold of equilibria contained in $\mathcal Z_{\mathcal C}$ and tangent to $\ker\nabla^2\mathcal H(x^*)$; hence $\mathcal Z_{\mathcal C}$ coincides locally with this orbit. The linearization of $\dot x=-\nabla\mathcal H(x)$ at $x^*$ is $-\nabla^2\mathcal H(x^*)$, whose transverse spectrum consists of $|\mathcal C|$ positive and $2N+M-|\mathcal C|-3$ negative eigenvalues, and it is the same along the orbit by the rigid-motion invariance of the closed loop.
\end{proof}

\begin{thm}\label{thm:almost_global}
	The set of initial states $x(0)\in\mathbb R^{2N+M}$ from  which the trajectory of \eqref{eq:proposed_controller} does not converge to $\mathcal Z_d$ has Lebesgue measure zero. Moreover, for every $p(0)$ except a set of measure zero, the set of initial phases $\omega(0)$ for which convergence to $\mathcal Z_d$ fails has measure zero. In either formulation, the corresponding trajectories satisfy $\operatorname{dist}(p(t),\mathcal S^*)\to0$.
\end{thm}
\begin{proof}
	By Theorems~\ref{thm:equilibrium_convergence} and \ref{thm:equilibrium_characterization}, every trajectory converges to a single point of $\mathcal Z_d$ or of $\mathcal Z_u:=\bigcup_{\mathcal C\neq\varnothing}\mathcal Z_{\mathcal C}$. By Theorem~\ref{thm:saddle}, the linearization at every $x^*\in\mathcal Z_u$ has at least one eigenvalue with positive real part. By the center-stable manifold theorem \cite{kelley1967stable}, $x^*$ admits a neighborhood $U_{x^*}$ such that the points whose forward trajectory remains in $U_{x^*}$ belong to a local center-stable manifold $W^{cs}_{\rm loc}(x^*)$ of dimension at most $2N+M-1$, which has measure zero. By Lindel\"of's lemma \cite[Thm.~30.3]{munkres2000topology}, $\mathcal Z_u$ is covered by countably many such neighborhoods $U_k$. If $x(t)\to x_\infty\in\mathcal Z_u$, then $x_\infty\in U_k$ for some $k$, and there exists $m\in\mathbb N$ such that $x(t)\in U_k$ for all $t\geq m$, i.e., $\Phi_m(x(0))\in W^{cs}_{{\rm loc},k}$, where $\Phi_t$ denotes the flow. Hence the set of such initial conditions is contained in $\bigcup_{k}\bigcup_{m\in\mathbb N}\Phi_{-m}(W^{cs}_{{\rm loc},k})$, a countable union of images of null sets under diffeomorphisms, which is a null set. The second statement follows from the Fubini--Tonelli theorem \cite[Thm.~2.37]{folland1999real}: integrating the indicator of this null set first over $\omega(0)\in\mathbb R^M$ and then over $p(0)\in\mathbb R^{2N}$ shows that its slice $\{\omega(0):x(0)\ \text{in the set}\}$ is null for almost every $p(0)$. The last statement follows from Proposition~\ref{prop:shape}.
\end{proof}

The second statement of Theorem~\ref{thm:almost_global} is relevant in practice: the initial phases are internal controller states chosen by the designer, and almost every choice is admissible regardless of the initial positions. A verifiable sufficient condition is obtained from the energy levels of the undesired equilibria.

\begin{corollary}\label{cor:desired_convergence}
	For every $x^*\in\mathcal Z_{\mathcal C}$, $\mathcal H(x^*)=k_p\sum_{\{i,j\}\in\mathcal C}(d_{ij}^*)^2$. Consequently, if $\mathcal E_I\neq\varnothing$ and
	\begin{equation}\label{eq:initial_energy_condition}
		\mathcal H(p(0),\omega(0))
		<
		k_p\min_{\{i,j\}\in\mathcal E_I}(d_{ij}^*)^2,
	\end{equation}
	then the trajectory converges to $\mathcal Z_d$.
\end{corollary}

\begin{proof}
	At $x^*\in\mathcal Z_{\mathcal C}$, $e_\omega=0$ and the only nonzero manifold errors are $\phi_{j|i}=\phi_{i|j}=-d_{ij}^*q_{j|i}$ for $\{i,j\}\in\mathcal C$, which gives the stated value. Since $\mathcal H$ is nonincreasing and the trajectory converges to an equilibrium $x_\infty$ with $\mathcal H(x_\infty)\leq\mathcal H(x(0))$, condition \eqref{eq:initial_energy_condition} excludes $x_\infty\in\mathcal Z_{\mathcal C}$ for every $\mathcal C\neq\varnothing$.
\end{proof}

\begin{corollary}\label{cor:star}
	If $\mathcal G$ is a star, then $\mathcal Z=\mathcal Z_d$, and every trajectory of \eqref{eq:proposed_controller} converges to $\mathcal Z_d$.
\end{corollary}
\begin{proof}
	A star has a single internal agent, so $\mathcal E_I=\varnothing$, and the claim follows from Theorems~\ref{thm:equilibrium_convergence} and \ref{thm:equilibrium_characterization}.
\end{proof}

\begin{corollary}\label{cor:connected}
	Let the links available for sensing and communication form a graph $\mathcal G_s=(\mathcal V,\mathcal E_s)$. If $\mathcal G_s$ is connected and $\mathcal G$ is any spanning tree of $\mathcal G_s$, then \eqref{eq:proposed_controller} implemented over $\mathcal G$ uses only links in $\mathcal E_s$, and Theorems~\ref{thm:equilibrium_convergence}--\ref{thm:almost_global} and Corollaries~\ref{cor:desired_convergence}--\ref{cor:star} hold. Conversely, connectivity of $\mathcal G_s$ is necessary for any formation controller that only uses links in $\mathcal E_s$.
\end{corollary}
\begin{proof}
	Every connected graph has a spanning tree, e.g., obtained by a breadth-first search, and the results above only require $\mathcal G$ to be a tree. If $\mathcal G_s$ is disconnected, the agents in one component receive no information about those in another, so their relative positions cannot be regulated.
\end{proof}

\begin{rem}[Phase initialization]\label{rem:init}
	Because the potential is quadratic in the lifted relative phases, an initial relative-phase error $e_{jk|i}(0)=2\pi+\epsilon$ is regulated to zero rather than to $2\pi$, which makes the agents complete an unnecessary turn around their host. This is avoided by initializing $\omega_{j|i}(0)$ with the current bearing of $p_j(0)-p_i(0)$ in the reference of host $i$, lifted so that $|e_{jk|i}(0)|\leq\pi$. Then every $\phi_{j|i}(0)$ is radial, and $\mathcal H(x(0))=\frac{k_p}{2}\sum_{\mathcal A}(\|p_j(0)-p_i(0)\|-d_{ij}^*)^2+\frac{k_\omega}{2}\|e_\omega(0)\|^2$.
\end{rem}

Undesired equilibria are thus entirely due to the internal edges, which are parameterized from both endpoints: a double-star tree ($|\mathcal E_I|=1$, $M=N$) has a single saddle family, with one negative, three zero, and $3N-4$ positive eigenvalues; a path has $|\mathcal E_I|=N-3$; and a star has none.

\subsection{Distributed Implementation in Local Frames}

In this subsection, we show that \eqref{eq:proposed_controller} can be implemented with each agent using only its own measurements, expressed in its own frame, and one scalar per arc sent from the host to the hosted agent; Fig.~\ref{fig5} summarizes the information involved.

Let $\mathcal O_k$ denote the local frame of agent $k$, assumed fixed in time, and let $F_k\in SO(2)$ denote its orientation with respect to an arbitrary reference frame used only for the analysis. Thus, all local frames share the same handedness, as in \cite{chen2020anglesss,liu2020switching}, which is necessary for any method without mirror-image ambiguity: a host with the opposite handedness would realize the mirrored angles of its star, and a hosted agent would misread the sign of $\delta_{j|i}$. Each host $i$ fixes the reference direction of its phases in its own body frame through an arbitrary constant rotation $B_i\in SO(2)$, i.e., the phase $\omega_{j|i}$ represents the physical direction $F_iB_iq(\omega_{j|i})$. Writing $F_iB_i=R(\vartheta_i)$, this direction equals $q(\omega_{j|i}+\vartheta_i)$. Since $\vartheta_i$ is constant and common to all phases hosted by $i$, it leaves the relative phases $e_{jk|i}$ unchanged; therefore, the analysis above holds verbatim with $\omega_{j|i}+\vartheta_i$ in place of $\omega_{j|i}$, and we can assume $\vartheta_i=0$ without loss of generality.

\emph{Host side.} Agent $i$ measures ${}^ir_{ij}:=F_i^\top(p_j-p_i)$ for every $j\in\mathcal N_i$ and computes
\begin{align*}
	{}^i\phi_{j|i}&=F_i^\top\phi_{j|i}={}^ir_{ij}-d_{ij}^*B_iq(\omega_{j|i}),\\{}^i\tau_{j|i}&=F_i^\top\tau_{j|i}=d_{ij}^*B_iEq(\omega_{j|i}),
\end{align*}
so that $({}^i\tau_{j|i})^\top({}^i\phi_{j|i})=\tau_{j|i}^\top\phi_{j|i}$. Moreover, $[A_\omega^\top e_\omega]_{(i,j)}$ depends only on the phases in $\mathcal A_i$. Hence, host $i$ integrates \eqref{eq:proposed_phase_controller} for all the phases it hosts without any communication, and the term $k_p\,{}^i\phi_{j|i}$ in its own input is computed without normalization.

\emph{Hosted side.} The hosted agent $j$ needs ${}^j\phi_{j|i}=F_j^\top\phi_{j|i}$. For $r_{ij}:=p_j-p_i\neq0$, let $\delta_{j|i}$ denote the signed angle from $r_{ij}$ to the direction $q_{j|i}$. It is a frame-independent scalar that host $i$ evaluates from ${}^ir_{ij}$ and $B_iq(\omega_{j|i})$ and transmits to agent $j$. Since $q_{j|i}=R(\delta_{j|i})r_{ij}/\|r_{ij}\|$ and planar rotations commute,
$${}^j\phi_{j|i}={}^jr_{ij}-d_{ij}^*R(\delta_{j|i})\frac{{}^jr_{ij}}{\|{}^jr_{ij}\|},$$
where ${}^jr_{ij}=F_j^\top(p_j-p_i)$ is measured by agent $j$ itself.

Accordingly, agent $k$ implements
\begin{equation}\label{eq:local_controller}
	{}^ku_k=-k_p\!\!\sum_{i:(i,k)\in\mathcal A}\!\!{}^k\phi_{k|i}+k_p\!\!\sum_{j:(k,j)\in\mathcal A}\!\!{}^k\phi_{j|k}=F_k^\top u_k,
\end{equation}
which represents exactly the same physical vector as \eqref{eq:proposed_position_controller}. Each arc requires a single scalar, $\delta_{j|i}$, sent from the host to the hosted agent; along internal edges, one scalar is sent in each direction. No vector quantity is exchanged, and no alignment between local frames is required.

\begin{rem}[Coincident neighbors]
	The phase dynamics and the host-side terms are smooth. The hosted-side reconstruction requires $r_{ij}\neq0$, i.e., it is undefined when a hosted agent coincides with its host. Such configurations include the collapsed edges of $\mathcal Z_{\mathcal C}$, which are unstable by Theorem~\ref{thm:saddle} but may be approached transiently. In practice, the last valid value of $\delta_{j|i}$ can be held, or the hosted-side term can be saturated, in a small neighborhood of $r_{ij}=0$.
\end{rem}

\section{Simulation}\label{sec:sim}
We illustrate the proposed method with two examples. In both, the prescribed configuration is a regular polygon with side $d^*$, i.e., $p^*_i=R\,q\big(2\pi(i-1)/N\big)$, $i=1,\ldots,N$, with circumradius $R=d^*/(2\sin(\pi/N))$, and the phases are initialized as in Remark~\ref{rem:init}.

\emph{Example (a): eight agents on a path.} Consider $N=8$ agents interacting over the path graph $\mathcal G$ with edges $\{i,i+1\}$, $i=1,\ldots,7$. Among the trees with eight vertices, the path has the largest number of internal edges, so it is the most demanding case for the equilibrium structure (Remark~\ref{rem:tree_choice}). Its internal agents are $2,\ldots,7$, which host the $M=12$ phases $\omega_{i-1|i}$ and $\omega_{i+1|i}$, $i=2,\ldots,7$. The $N-2=6$ relative-phase relations are $e_{(i-1)(i+1)|i}$, $i=2,\ldots,7$, and the $|\mathcal E_I|=5$ internal edges give rise to $2^5-1=31$ families of undesired equilibria. The prescribed configuration is a regular octagon with side $d^*=10$\,m, i.e., $R\approx13.07$\,m, so that $\Delta^*_{(i-1)(i+1)|i}=135^\circ$ for $i=2,\ldots,7$, as described in Section~\ref{s3}. The controller gains are $k_p=1$ and $k_\omega=100$, and the initial positions are drawn uniformly from $[-20,20]^2$\,m.

\emph{Example (b): twenty agents on a binary tree.} To put the method to a larger test, consider $N=20$ agents interacting over a balanced binary tree, in which agent $i$ is linked to agent $\lfloor i/2\rfloor$, $i=2,\ldots,20$, and forming a regular $20$-gon with side $d^*=5$\,m, i.e., $R\approx15.98$\,m. Most links of this tree are chords of the polygon, with prescribed lengths between $5$ and $32$\,m, so the interaction graph is unrelated to the geometry of the target shape. The internal agents $1,\ldots,10$ host $M=28$ phases with $N-2=18$ relative-phase relations, whose desired values follow from $p^*$ as described in Section~\ref{s3}, and the $|\mathcal E_I|=9$ internal edges give rise to $2^9-1=511$ families of undesired equilibria. Since the slowest local convergence rate decreases as the formation grows, we use the larger gains $k_p=10$ and $k_\omega=1000$, which yield a local convergence rate similar to that of example (a). The initial positions are drawn uniformly from $[-25,25]^2$\,m.

\begin{figure}
	\centering
	\includegraphics[width=0.86\columnwidth]{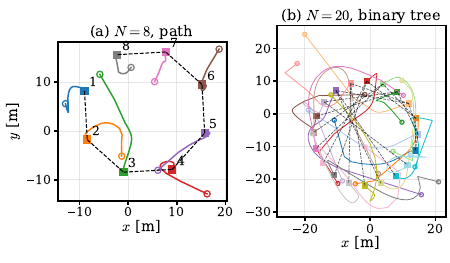}
	\caption{Trajectories of the agents in the two examples. (a) Eight agents on the path graph forming a regular octagon with side $10$\,m. (b) Twenty agents on a balanced binary tree forming a regular $20$-gon with side $5$\,m. Circles and squares denote the initial and final positions, respectively, and dashed lines indicate the links of the tree; agent labels are omitted in (b) for readability.}
	\label{fig:tree_trajectory}
\end{figure}

\begin{figure}
	\centering
	\includegraphics[width=\columnwidth]{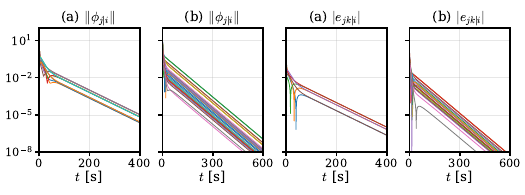}
	\caption{Manifold errors $\|\phi_{j|i}\|$ [m] and relative-phase errors $|e_{jk|i}|$ [rad] in the two examples, in logarithmic scale: (a) twelve arcs and six relations; (b) twenty-eight arcs and eighteen relations.}
	\label{fig:tree_errors}
\end{figure}

Fig.~\ref{fig:tree_trajectory} shows the trajectories of the agents in both examples. The agents converge to the prescribed regular polygons using only the $N-1$ links of the corresponding trees, with an arbitrary final position and orientation. Since the proposed controller does not include collision avoidance, the trajectories may cross during the transient. Fig.~\ref{fig:tree_errors} shows the manifold errors $\|\phi_{j|i}\|$ and the relative-phase errors $|e_{jk|i}|$ in logarithmic scale; their linear decay after the initial transient is consistent with the local exponential stability of Proposition~\ref{prop:local_exp}. Note that the initial energies, $\mathcal H(x(0))=2482.5$ in example (a) and $\mathcal H(x(0))=58153.5$ in example (b), largely exceed the corresponding energy levels $k_p\min_{\mathcal E_I}(d^*_{ij})^2=100$ and $250$ of Corollary~\ref{cor:desired_convergence}, so convergence in these cases is guaranteed by Theorem~\ref{thm:almost_global} rather than by the energy condition.

To assess the almost-global result, we repeated both simulations from random initial conditions, with phases drawn uniformly from $[-\pi,\pi]$ and without the initialization of Remark~\ref{rem:init}: $300$ runs for example (a), with positions drawn uniformly from $[-25,25]^2$\,m, and $100$ runs for example (b), with positions drawn uniformly from $[-30,30]^2$\,m. All trajectories converged to $\mathcal S^*$, with largest residual shape errors of $1.2\times10^{-8}$\,m after $1500$\,s in example (a) and $7.9\times10^{-10}$\,m after $3000$\,s in example (b), measured as the maximum agent-wise distance after the best alignment by a planar rigid motion.

\section{Conclusion}\label{sec:conclusion}
This paper developed a formation controller based on parameterized circular manifolds and virtual-phase coordination that achieves a prescribed planar formation, up to translation and rotation, on any connected graph by operating on one of its spanning trees. The controller is a gradient flow whose undesired equilibria are the collapses of subsets of internal tree edges, each with as many unstable directions as collapsed edges; hence, the desired formation is locally exponentially stable and almost globally attractive, and globally attractive for star trees. It is implemented in mutually misaligned local frames with one scalar per arc.

Future work will exploit the links outside the selected tree, e.g., for redundancy, which requires handling the redundant constraints of cycles; the tangential component of the guiding vector field, set to zero here, to induce coordinated motions through the common evolution of the virtual phases; and three-dimensional formations and nonholonomic vehicles with constant speeds.
\bibliographystyle{IEEEtran}

\bibliography{Control_Refer_Logic_New,connected_extra}

\end{document}